\documentclass[lettersize,journal]{IEEEtran}
\usepackage{amsmath,amsfonts}
\usepackage{algorithmic}
\usepackage{algorithm}

\floatname{algorithm}{Algorithm}

\usepackage{array}
\usepackage[caption=false,font=normalsize,labelfont=sf,textfont=sf]{subfig}
\usepackage{textcomp}
\usepackage{stfloats}
\usepackage{url}
\usepackage{verbatim}
\usepackage{graphicx}
\usepackage{cite}
\usepackage{gensymb}
\usepackage{xcolor}
\usepackage{color}
\usepackage{colortbl}
\usepackage{adjustbox}
\usepackage{multirow}
\usepackage{array} 
\usepackage{tabularray}
\usepackage{balance}
\usepackage{hhline}
\usepackage{makecell} 
\usepackage{enumerate}
\usepackage{threeparttable}

\begin{document}

\title{A Measurement-Based Parameterization of Physics Reflection Models for Terahertz Communication}
\author{Taihao Zhang,~\IEEEmembership{Member,~IEEE}, Chenzhou Lin, Cunhua Pan,~\IEEEmembership{Senior Member,~IEEE}, Hong Ren,~\IEEEmembership{Member,~IEEE}, Ruyi Liu, Yongchao He, Tian Qiu, Bingchang Hua, Jiangzhou Wang,~\IEEEmembership{Fellow,~IEEE}
\thanks{This work was supported by the Key Research and Development Projects under Grant 2023YFB2905100. T. Zhang, C. Lin, C. Pan, H. Ren, R. Liu, Y. He, T. Qiu and J. Wang are with the National Mobile Communications Research Laboratory, School of Information Science and Engineering, Southeast University, Nanjing 211189, China (e-mail: {taihao, 213223687, cpan, hren, 213241382, heyongchao, tianqiu, j.z.wang}@seu.edu.cn). B. Hua is with the Purple Mountain Laboratories, Nanjing, Jiangsu 211111, China (e-mail: huabingchang@pmlabs.com.cn)}
}

\maketitle

\begin{abstract}
The accurate modeling of reflection coefficients is pivotal for developing reliable channel models in emerging terahertz (THz) communications.
This study establishes a 300$\sim$400 GHz channel measurement platform to measure the reflection coefficients of various materials. Based on the analysis of measured data, we propose the single-layer interference with an extended-parameterized Lorentz/Drude (SLI-EPLD) reflection coefficient model. In this model, a sub-band modeling strategy is adopted to characterize the variation of reflection coefficients with frequency, while a parameterized mapping approach is employed to ensure the stability of model parameters. Furthermore, the weighted sub-band fitting for trend regression (WF-TREND) algorithm is introduced to achieve precise sub-band parameter fitting. Validation  results demonstrate superior performance to existing models across multiple materials. The reflection coefficient model established in this work serves as a critical foundation for channel modeling in 300$\sim$400 GHz for high-THz communication.
\end{abstract}


\section{Introduction}
The terahertz (THz) band (0.1$\sim$10 THz), recognized for its vast spectrum resources, is widely regarded as a key enabler for 6G wireless communications to achieve terabit-per-second (Tbps) data rates and immersive applications \cite{thzbu,thz1,thzbu2}. Accurate channel modeling serves as the cornerstone for THz system design. This is particularly critical in complex environments, where reflection paths play an important role in non-line-of-sight (NLOS) propagation in the THz band \cite{NLOS}. Furthermore, precise characterization of reflection coefficients is essential for completing high-frequency reflection databases in ray-tracing (RT) software \cite{RT}.
Therefore, accurate modeling of material reflection coefficients in the THz band holds significant theoretical and practical importance.

Recent efforts have advanced the characterization of reflection coefficients in the THz band.
\cite{D} conducted measurements on indoor materials in 110$\sim$170 GHz, qualitatively analyzing angle and frequency dependence of reflection coefficients. 
\cite{bizhi} performed 219$\sim$224 GHz narrowband measurements on building materials using a frequency-domain platform, proposing a modified single-layer interference (SLI) model. \cite{Chang2023} employed a time-domain platform to measure diverse materials in 220$\sim$320 GHz, establishing a frequency–angle two-dimensional reflection coefficient (FARC) model. However, all models fail to explicitly account for the frequency dispersion of their parameters, meaning the frequency-dependent variation of material properties, which is a critical effect and cannot be neglected in wideband THz scenarios.
Moreover, the International Telecommunication Union (ITU) updated its reflection coefficient guidelines in Rec. ITU-R P.2040-4 \cite{ITU}. Although the ITU model incorporates frequency-dependent parameters, its reliance on a simple, few-parameter, and macroscopic empirical formulation may limit its capability to capture complex dispersion behaviors.
Thus, two critical research gaps remain: ($i$) a scarcity of experimental reflection coefficient data for diverse materials in high-frequency bands, and ($ii$) the insufficient capability of existing models to accurately characterize the properties of reflection coefficient.

The main contributions of this paper is summarized as follows:
($i$) We conduct channel measurements across materials in 300$\sim$400 GHz, supplementing the reflection coefficient dataset for higher frequency bands.
($ii$) We propose the single-layer interference with an extended-parameterized Lorentz/Drude (SLI-EPLD) model, 
which employs a parameterized mapping to ensure parameter stability and effectively captures reflection characteristics, especially frequency dispersion.
($iii$) We develop the weighted sub-band fitting for trend regression
(WF-TREND) algorithm, specifically tailored to reconstruct parameter variations with frequency.


\section{The Measurement Campaign for reflection coefficients in the THz Band} \label{II}
\subsection{Measurement Setup in 300$\sim$400 GHz}
To support reflection coefficient modeling, we employ a VNA-based frequency-domain channel measurement system. The detailed principles can be referenced in \cite{taihao}. As Fig. \ref{figreal}, in the transmitter module, the VNA generates an RF signal in 11.11$\sim$14.81 GHz, which is up-converted by a $\times$ 27 frequency multiplier (FM). 
Concurrently, a local oscillator (LO) outputs a signal in 12.50$\sim$16.66 GHz, which is up-converted by a $\times$ 24 FM. The two resulting signals are mixed to generate a 279 MHz intermediate frequency (IF) reference signal. An analogous process occurs in the receiver module to produce the corresponding test signal.
The VNA measures the complex ratio between the test and reference signals to extract the amplitude and phase information of $S_{21}$. The key configuration parameters and setup are listed in Table \ref{parameters}.
\begin{table}[h]
	\centering
	\caption{Configuration parameters of the Measurement System} \label{parameters}
    \resizebox{1.0\linewidth}{!}{ 
        \begin{tabular}{c|c|c|c}
		\hline		 \rowcolor{gray!25}Parameter & Value&Parameter & Value\\
		\hline
            Start frequency & 300 GHz&End frequency & 400 GHz\\
            \rowcolor{gray!10}Sweeping points & 12001 &Sweeping interval & 11.67 MHz\\
            Antenna gain at Tx/Rx & 25 dBi&HPBW of Tx/Rx&E 7.7$\degree$, H 8.5$\degree$\\
            \rowcolor{gray!10}Average noise floor & -145 dBm &Test power & 0.5 mW \\
            Incident angle&10$\degree$:10$\degree$:80$\degree$&Measurement radius&0.40 m\\
            \rowcolor{gray!10}Temperature&22$\sim$25$\degree \mathrm{C}$ & humidity & 30$\%$ $\sim$ 40$\%$\\
            \hline
	\end{tabular}
        }
\end{table}
\subsection{Measurement Procedures}
\begin{figure}
	\centering
	\includegraphics[width=6.5cm]{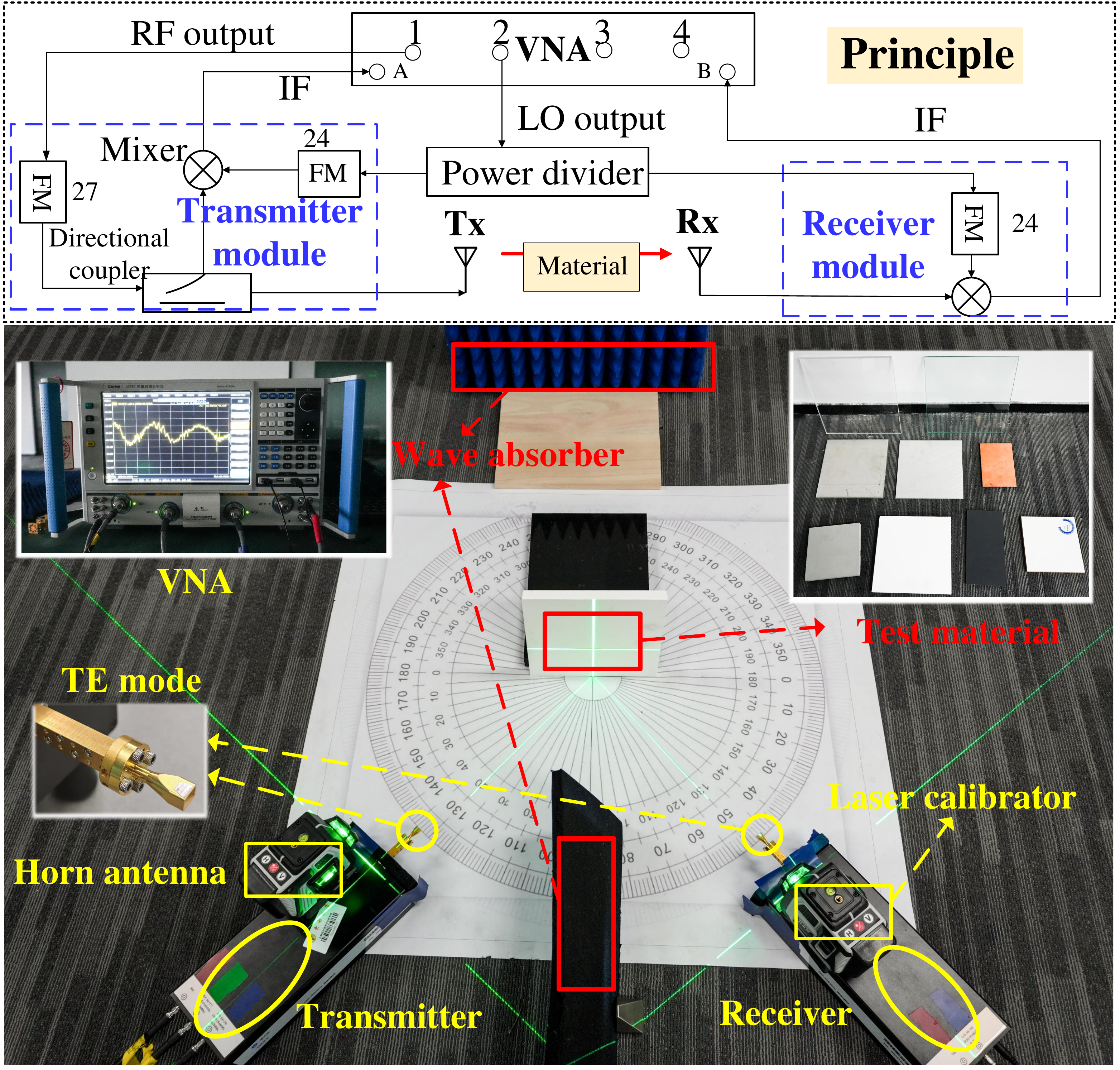}
	\caption{Measurement principle and setup of the reflection coefficient.}
	\label{figreal} 
\end{figure}
Fig. \ref{figreal} illustrates the practical measurement setup and environment. Prior to formal measurements, a direct-link calibration is performed to eliminate the influence from the frequency response introduced by the RF antennas, the VNA, and the connecting cables. Two laser alignment tools are used to ensure angle accuracy. 
The measurement radius is set to 40 cm to satisfy the far-field condition\footnote{With the maximum horn antenna aperture of 1 cm, the maximum Rayleigh distance for the measured frequency band is approximately 26.67 cm.}, which ensures the accuracy of angle measurement under the plane-wave assumption. And wave absorbers are placed to prevent interference with the received signal.
The incident angle ranges from 10$\degree$ to 80$\degree$ with step of 10$\degree$.
The measurements are performed for the transverse electric (TE) mode.
A total of ten materials are tested. The dimensions of each material sample are large enough to prevent edge diffraction effects during measurements.

The collected data reveal that the reflection from the copper (Cu) metal is the strongest and most stable\footnote{Cu serves as an ideal reference due to its stability and strong reflection among the tested materials, as evidenced in Fig. \ref{max}. The thin natural oxide layer is negligible, and the ratio method further cancels any residual effects.}. Therefore, the metal reference ratio method is employed to calculate the reflection coefficient, given by
\begin{align}
\varGamma _{\mathrm{mea}}=\left| \frac{S_{21,\mathrm{Material}}}{S_{21,\mathrm{Cu}}} \right|,
\end{align}
where $S_{21,\mathrm{Cu}}$ and $S_{21,\mathrm{Material}}$ are the linear-scale magnitudes of the measured $S_{21}$ parameters for the copper reference and the tested materials, respectively.
This ratio method is robust against system variations such as path loss and antenna patterns, making it suitable for THz band measurements with high-gain directional antennas.

\section{The Thz-band Reflection Characterization and Analysis}\label{III}
\subsection{Measurement Results and Analysis}\label{meas}
\begin{figure}
	\centering
	\includegraphics[width=8cm]{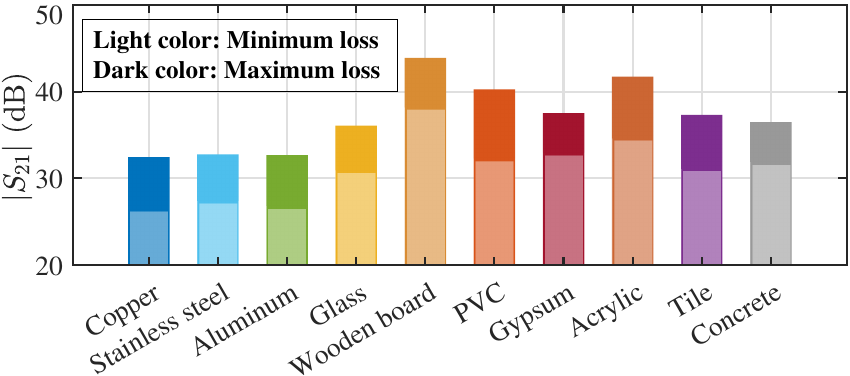}
	\caption{The $\left| S_{21} \right|$ at an incident angle of 30$\degree$ in 300$\sim$400 GHz.}
	\label{max} 
\end{figure}
Based on the measured data, we compare $\left| S_{21} \right|$ of different materials, as illustrated in Fig. \ref{max}. It can be observed that the reflection losses of metals are roughly the same and lower than those of non-metals, whereas the reflection loss of non-metals vary significantly.

\begin{figure}
	\centering  \includegraphics[width=8cm]{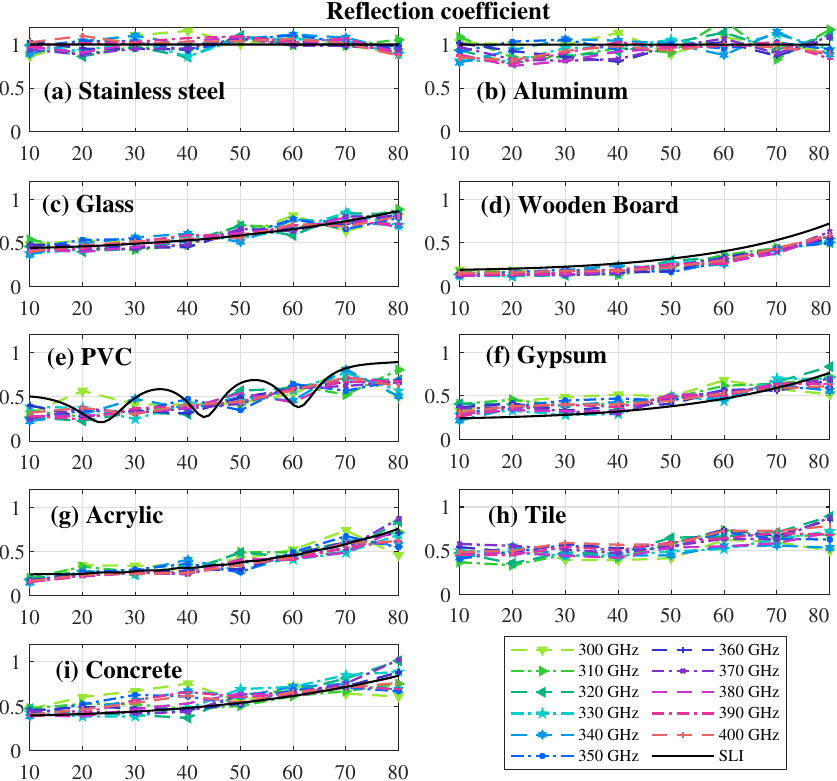}
	\caption{Measured and SLI-modeled reflection coefficients of various materials as a function of incident angle (deg) at different frequencies.}
	\label{angle} 
\end{figure}

Fig. \ref{angle} illustrates the angle variation of the measured reflection coefficient for each material at different frequencies. First, the reflection coefficients of non-metal materials increase with the incident angle, whereas those of metal materials show relatively weak angle dependence. Second, at a fixed angle, materials such as stainless steel, glass and wooden board exhibit slight variations in reflection coefficient across frequency, while the remaining materials demonstrate considerably stronger frequency fluctuations. 

\begin{figure}
	\centering
	\includegraphics[width=6cm]{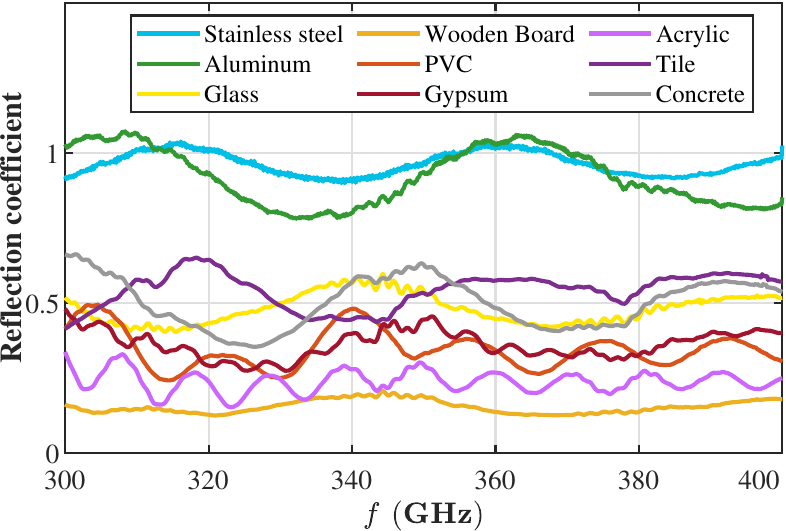}
	\caption{Measured reflection coefficients of various materials as a function of frequencies at an incident angle of 30$\degree$.}
	\label{f} 
\end{figure}

To further investigate the frequency dependence of the reflection coefficients, we plot their variation against frequency in 300$\sim$400 GHz at an incident angle of 30$\degree$, as shown in Fig. \ref{f}. All materials exhibit quasi-periodic fluctuations in their reflection coefficients across the frequency range. Metal materials show minimal fluctuations\footnote{The oxide layer on the surface of the metal samples results in fluctuations of the reflection coefficients around 1.}, whereas non-metal materials exhibit significantly different fluctuation periods.
\subsection{Comparison with the Classic Model}
Based on the measurement results and preliminary analysis in Section \ref{meas}, it is observed that the reflection coefficients exhibit angle and frequency dependence. This characteristic is classically represented by the SLI model. 
Its formulation for the TE mode is given by
\begin{align}\label{ITU}
	\varGamma =\left| \frac{R\left( 1-e^{-j2q} \right)}{1-R^2e^{-j2q}} \right|, R=\frac{\cos \theta -\sqrt{\eta -\sin ^2\theta }}{\cos \theta +\sqrt{\eta -\sin ^2\theta }},
\end{align}
with $q=\frac{2\pi d}{\lambda}\sqrt{\eta -\sin \theta ^{2}}$,
where $\theta $ is the incident angle, $d$ is the thickness of the material, $\lambda$ is the wavelength.
Frequency dependence lies in the complex permittivity $\eta$, which employs a classical macroscopic empirical formula as 
\begin{align}\label{ituyi}
\eta =\varepsilon _r-j\frac{\sigma}{2\pi f\varepsilon _0},
\end{align}
where $\varepsilon _r$ is the real part of the complex permittivity, $\sigma$ is the conductivity and $\varepsilon _0$ is the permittivity of free space. 
The SLI model is plotted as a black curve in Fig. \ref{angle}, and the parameters $\left\{ \varepsilon _r,\sigma \right\} $ are taken from Refs. \cite{bizhi}\cite{ITU}.


However, there are discrepancies between the SLI model and the measured data. These can be attributed to the absence of data in specific high-frequency bands, and the model's insufficient consideration of the frequency dependence of material properties.
Furthermore,  the model's description of the complex permittivity adopts a macroscopic empirical formula \eqref{ituyi} with relatively few parameters, whose descriptive capability may be limited.
Therefore, we propose a novel model to more accurately characterize  the frequency dispersion properties in the THz band, thereby enabling more precise prediction of reflection coefficients.

\section{Reflection Coefficient Modeling for the THz Band}\label{IIII}
To more accurately characterize the reflection properties of materials in the THz band, we propose the SLI-EPLD model. Outperforming the traditional SLI model, this model incorporates a surface roughness correction factor and an extended and parameterized Lorentz/Drude model, enabling an accurate description of its angle and frequency dependence.

\subsection{Model Construction}\label{pan}
\subsubsection{Model Formulation}\label{model}
Based on the SLI model, the fundamental structure of the proposed model is given by
\begin{align}\label{tao}
\varGamma \left( f,\theta \right) =\rho \left| \frac{R\left( f,\theta \right) \left( 1-e^{-j2q\left( f,\theta \right)} \right)}{1-R^2\left( f,\theta \right) e^{-j2q\left( f,\theta \right)}} \right|.
\end{align}
Since the surface roughness of materials is non-negligible, the reflection coefficient must be multiplied by a scattering loss factor $\rho$, which attenuates the reflected intensity \cite{Rap02}. The factor $\rho$ is given by \cite{Boi87}
\begin{align}\label{rho}
\rho =e^{-8\left( \frac{\pi \sigma _h\cos \theta }{c}f \right) ^2}I_0\left[ 8\left( \frac{\pi \sigma _h\cos \theta }{c}f \right) ^2 \right],
\end{align}
where $\sigma _h$ is the root mean square (RMS) of the surface roughness and $I_0\left[  \,\,\right]$ is the zeroth-order Bessel function of the first kind.

Then, the angle and frequency dependence are captured by the parameters $R\left( f,\theta \right)$ and $q\left( f,\theta \right)$, given by
\begin{align}\label{R}
R\left( f,\theta \right) =\frac{\cos \theta -\sqrt{\eta \left( f \right) -\sin ^2\theta }}{\cos \theta +\sqrt{\eta \left( f \right) -\sin ^2\theta }} ,
\end{align}
\begin{align}
q\left( f,\theta \right) =\frac{2\pi d}{\lambda}\sqrt{\eta \left( f,\theta \right) -\sin^{2} \theta }.
\end{align}
For the complex permittivity, the SLI-EPLD model adopts more fundamental and microscopic Lorentz/Drude structure. For non-metal, the Lorentz model is employed, given by
\begin{align}\label{L}
\eta _{\mathrm{L}}\left( f \right) =1+\frac{w_{\mathrm{p}}^{2}}{w_{0}^{2}-\left( 2\pi f \right) ^2-2\pi j\gamma f},
\end{align}
where $w_{\mathrm{p}}$ denotes the plasma frequency, $w_{0}$ is the resonant frequency.
Correspondingly, for metal, the Drude model is used, given by
\begin{align}\label{D}
\eta _{\mathrm{D}}\left( f \right) =1-\frac{w_{\mathrm{p}}^{2}}{\left( 2\pi f \right) ^2+2\pi j\gamma f}.
\end{align}
In \eqref{L} and \eqref{D}, $\gamma$ represents the damping coefficient which is inversely proportional to the relaxation time $\tau $, i.e. $\gamma =\frac{1}{\tau}$. 

\subsubsection{Parameterization Strategy}
If the physical parameters from Section \ref{pan} \textit{1)} are directly treated as unknowns for estimation, numerical instability may arise due to parameters having excessively large or small numerical magnitudes. To address this, SLI-EPLD introduces a layer of parameterization, mapping the physically meaningful parameters $\left\{ \sigma _h,w_{\mathrm{p}}^{2},w_{0}^{2},\gamma \right\}$ to a set of macroscopic parameters $\mathbf{P}=\left[ p_1,p_2,p_3,p_4 \right] ^{\mathrm{T}}$, which exhibit more stable numerical ranges.

Specifically, with all frequency $f$ expressed in GHz. In \eqref{rho}, $\frac{\pi ^2\sigma _{h}^{2}}{c^2}$ has a magnitude on the order of $10^{-15}$. Direct estimation of such extreme values is numerically unstable. Therefore, the parameterization is introduced as
\begin{align}
\rho _s=e^{-p_1f^2\cos ^2\theta }I_0\left[ p_1f^2\cos ^2\theta \right] , \,\, \lg p_1=b_{1}.
\end{align}
Moreover, the similar parameterization is applied to \eqref{L} and \eqref{D} after dividing both the numerator and the denominator by $2\pi \gamma $, yielding
\begin{align}
\eta _{\mathrm{nm}}\left( f \right) =1+\frac{p_2}{p_3-p_4f^2-jf},\,
\eta _{\mathrm{m}}\left( f \right) =1-\frac{p_2}{p_4+jf}.
\end{align}
As emphasized, due to the fact that the relaxation time parameters $\frac{1}{\gamma}$ exhibit variation with frequency\footnote{It reveals a continuous distribution of relaxation times within the material \cite{PLANK}. Measurements at different frequencies effectively probe different weighted portions of this distribution. Consequently, a more accurate model should account for the influence of $f$ on $\gamma$.}\cite{COFFEY}, $\left\{ p_2,p_3,p_4 \right\}$ must be estimated separately for different frequencies.
Hence, we employ log-linear regression analysis, yielding
\begin{align}\label{LOG}
\lg p_l=k_{l}f+b_{l},l=1,...,4,k_{1}=0.
\end{align}
Finally, our proposed SLI-EPLD model is,
\begin{align}
\varGamma \left( f,\theta \right) =\rho _s\left| \frac{R\left( \eta _{\mathrm{nm}/\mathrm{m}}\left( f \right) ,\theta \right) \left( 1-e^{-j2q\left( \eta _{\mathrm{nm}/\mathrm{m}}\left( f \right) ,\theta \right)} \right)}{1-R^2\left( \eta _{\mathrm{nm}/\mathrm{m}}\left( f \right) ,\, \theta \right) e^{-j2q\left( \eta _{\mathrm{nm}/\mathrm{m}}\left( f \right) ,\theta \right)}} \right|.
\end{align}

\subsection{WF-TREND Algorithm}
\begin{algorithm}[t]
	\caption{WF-TREND}
	\label{alg1}
	\renewcommand{\algorithmicrequire}{\textbf{Input:}}
	\renewcommand{\algorithmicensure}{\textbf{Output:}}
	\begin{algorithmic}[1]
		\REQUIRE 
		$\left[ \boldsymbol{f},\boldsymbol{\theta },\mathbf{\Gamma }_{\mathrm{meas}}\left( \boldsymbol{f},\boldsymbol{\theta } \right) \right] \in \mathbb{R} ^{N_{\mathrm{train}}\times 3}$, $d$, $\Delta f$ and $W$.

        \STATE The number of sub-bands: $I=\lceil \left( f_{\mathrm{end}}-f_{\mathrm{start}} \right) /\Delta f \rceil$.

        \STATE \textbf{Global initialization}: $\mathbf{P}_{_{0}}^{\mathrm{int}}$
        
        \FOR{$i=1$ to $I$}
        \STATE \textbf{Local initialization} (Step 5$\sim$7) :
        \STATE  The starting of sliding windows: $W_s=\max \left\{ 0,i-W \right\}$
        \STATE Calculate $\epsilon _j=\left\| \mathbf{\Gamma }_{\mathrm{model}}\left( \boldsymbol{f},\boldsymbol{\theta };\mathbf{P}_{j}^{\mathrm{int}} \right) -\mathbf{\Gamma }_{\mathrm{meas}} \right\| _{2}^{2}$ and \\$\bar{\kappa}_j=\left( \frac{1}{\epsilon _j} \right) /\sum_{j=W_s}^{i-1}{\left( \frac{1}{\epsilon _j} \right)}$ for $W_s\leq j\leq i-1$.

        \STATE Calculate the quality-weighted initial parameters for the $i$-th band: $\mathbf{P}_{i}^{\mathrm{int}}=\sum_{j=W_s}^{i-1}{\bar{\kappa}_j\mathbf{P}_j}$.
        \STATE 
        Use the LM algorithm to solve Problem \eqref{cv} to obtain $\mathbf{P}_i$ and root mean square error (RMSE) $\boldsymbol{\epsilon }_i$.
        \ENDFOR
        \STATE Obtain the set of sub-band parameters $\left\{ \mathbf{P}_i \right\} _{i=1}^{I}$, 
        \STATE Perform log-linear regression on the set $\left\{ \mathbf{P}_i \right\} _{i=1}^{I}$ according to \eqref{LOG} to obtain the final trend parameters.
		\ENSURE $\mathbf{K}=\left\{ k_{l}|l=1,...,4 \right\}$ and $\mathbf{B}=\left\{ b_{l}|l=1,...,4 \right\}$.   
	\end{algorithmic}
\end{algorithm}
To precisely characterize the frequency dependence of the reflection coefficient model, we propose a novel WF-TREND algorithm, which aims to robustly extract the macroscopic trend parameters of materials reflection from wide band measurement data. The objective function is as follows,
\begin{align}\label{cv}
\mathbf{P}=arg\min_{\mathbf{P}} \left\| \mathbf{\Gamma }_{\mathrm{model}}\left( \boldsymbol{f},\boldsymbol{\theta };\mathbf{P} \right) -\mathbf{\Gamma }_{\mathrm{meas}} \right\| _{2}^{2},
\end{align}
where $\boldsymbol{f}$ and $\boldsymbol{\theta}$ are the vectors of frequency and incident angle samples, respectively.

The detailed algorithm is given in Algorithm \ref{alg1}.
Its inputs are the measured reflection coefficient $\left[ \boldsymbol{f},\boldsymbol{\theta },\mathbf{\Gamma }_{\mathrm{meas}}\left( \boldsymbol{f},\boldsymbol{\theta } \right) \right] \in \mathbb{R} ^{N_{\mathrm{train}}\times 3}$ ($N_{\mathrm{train}}=N_{\mathrm{all}}\times 60\%$ is the size of the training set, selected via stratified random sampling to ensure uniform coverage of frequency and angle), material thickness $d$, division bandwidth  $\Delta f$ and sliding window length $W$. Given the nonconvex and nonlinear nature of the objective function, the Levenberg-Marquardt (LM) algorithm is well-suited for this problem due to its robust convergence properties \cite{Fischer2024}.

The core of the WF-TREND algorithm lies its two-layer architecture. The outer layer sub-band fitting separates the wide band problem into manageable narrow-band sub-problems to capture frequency-dependence, which are usually smoothed out in a global fitting process. Then, the inner layer trend regression unifies these sub-results to build a consistent macroscopic model of the entire band.
Fitted with the WF-TREND, the SLI-EPLD model parameters are given in Table \ref{tab2}.

\subsection{Performance Evaluation and 
Comparative Analysis}
\begin{table*}[t]
\centering
\caption{Fitted Model Parameters and Performance Comparison.}
\label{tab2}
\renewcommand\arraystretch{1} 
\adjustbox{max width=\textwidth}{
\begin{threeparttable}
\begin{tabular}{c|c|c|c|c|c|c|c!{\vrule width 1.3pt}c|c|c|c} 
\hline
\rowcolor{gray!25}\multicolumn{1}{c|}{} &  $ p_1 $  & \multicolumn{2}{c|}{ $ p_2 $ } & \multicolumn{2}{c|}{ $ p_3 $ } & \multicolumn{2}{c!{\vrule width 1.2pt}}{ $ p_4 $ } &  \multicolumn{4}{c}{ RMSE\tnote{*} }\\
\hiderowcolors
\noalign{\vskip -0.45pt} 
\hhline{~-----------}
\rowcolor{gray!25}\multicolumn{1}{c|}{\multirow{-2}{*}{Material}} &  $ b_1 $  &  $ k_2 $  &  $ b_2 $  &  $ k_3 $  &  $ b_3 $  &  $ k_4 $  &  $ b_4 $  & SLI-EPLD  & Ref. \cite{bizhi}&Ref. \cite{Chang2023} &Ref. \cite{ITU}\\
\hiderowcolors
\hline 
Glass & -14.7072 & -0.1444 & 2.9835 & 0.0767 & 3.0687 & 0.0684 & -2.4791 & 0.0050 & 0.0247& 0.0487 & 0.0085\\
\rowcolor{gray!10}Wooden Board & -14.3572 & -0.2067 & 2.8802 & 0.0969 & 3.0588 & 0.0655 & -2.4679 & 0.0047 & 0.0300& 0.0219&0.0066\\
PVC & -14.5305 & -0.1218 & 2.8488 & 0.1210 & 2.9848 & 0.1042 & -2.5683 & 0.0034 & 0.0640& 0.0708&0.0295\\
\rowcolor{gray!10}Gypsum & -14.0106 & -0.2054 & 3.1136 & 0.0930 & 3.0711 & 0.1016 & -2.5586 & 0.0062 & 0.0453&0.0509 &0.0254\\
Acrylic & -14.4616 & -0.2664 & 3.1406 & 0.1188 & 2.9599 & 0.1855 & -2.7820 & 0.0076 & 0.0351&0.0388 &0.0145\\
\rowcolor{gray!10}Tile & -14.6197 & -0.0943 & 2.5385 & 0.1776 & 2.7872 & 0.1444 & -2.5832 & 0.0057 & 0.0477& 0.0586& 0.0101\\
Concrete & -13.9350 & -0.0710 & 2.4141 & 0.2143 & 2.6463 & 0.1726 & -2.6662 & 0.0078 & 0.0832& 0.0848& 0.0299\\
\rowcolor{gray!10}Aluminum & -14.8545 & - & 4.3966 & - & - & - & -1.0000 & 0.0018 & 0.0251& 0.0198 &0.0136\\
Stainless steel & -15.0567 & - & 4.4962 & - & - & - & -1.0056 & 0.0013 &0.0171 &0.0214 &0.0124\\
\hline
\end{tabular}

 \begin{tablenotes}
        \footnotesize
        \item[*]For fair comparison, all models are fitted to our measurement data using the WF-TREND. The same procedure is applied in Table \ref{error}. 
      \end{tablenotes}

\end{threeparttable}
}
\end{table*}

\begin{figure}[t]
	\centering
	\includegraphics[width=8.5cm]{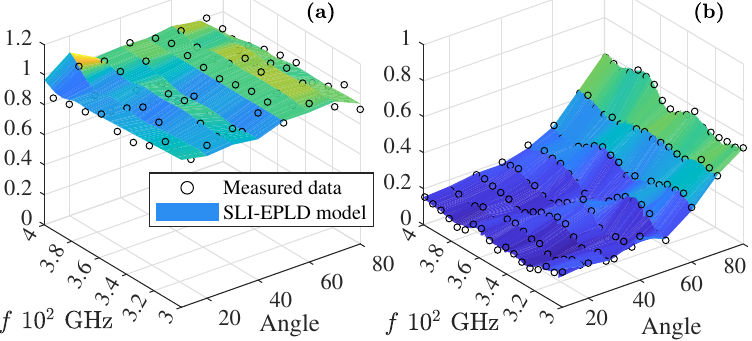}
	\caption{Stainless Steel (a) and Wooden Board (b) Reflection Coefficient: SLI-EPLD Model Fit versus Measurement over a range of Angles and Frequencies.}
	\label{fit1} 
\end{figure}

To visually demonstrate the model's validity, the measured and modeled reflection coefficients for stainless steel (metal) and wooden board (non-metal) across frequencies and angles in 300$\sim$400 GHz are compared in Fig. \ref{fit1}. The results indicate that the SLI-EPLD model accurately characterizes the angle and frequency dependence of the reflection coefficients.

\begin{figure}
	\centering
	\includegraphics[width=8.5cm]{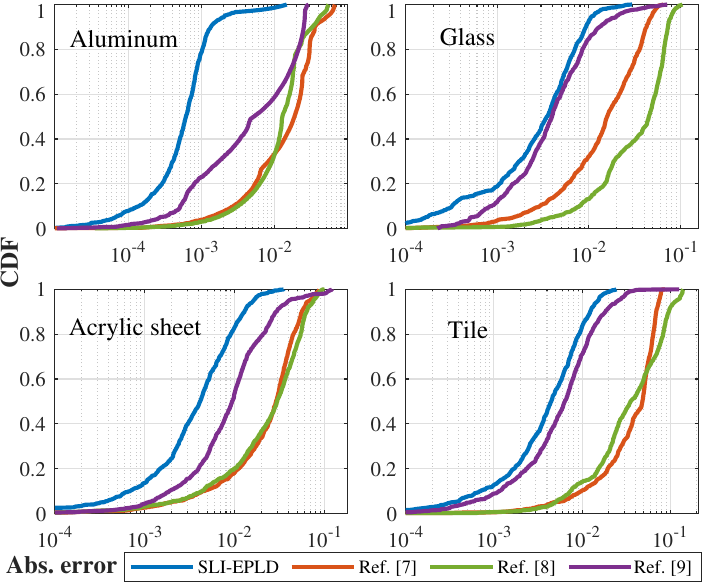}
	\caption{CDFs of the absolute error at the test set for four materials.}
	\label{abs} 
\end{figure}

To quantify the accuracy of the SLI-EPLD model, we calculate its RMSE and compare it with other models. As summarized in Table \ref{tab2}, the proposed model performs better. Furthermore, the absolute error cumulative distribution functions (CDFs) for all models are evaluated using the test dataset. The corresponding error value at the 90$\%$ confidence level is listed in Table \ref{error}. Taking Fig. \ref{abs} as an example, it compares the CDFs for four materials under each model. The proposed model's CDF rises the fastest, indicating smaller, more concentrated errors.

\begin{table}
	\centering
	\caption{The 90 $\%$ confidence bound on the abs. error at the Test Set .} \label{error}
    \resizebox{1.0\linewidth}{!}{ 
        \begin{tabular}{c|c|c|c|c}
		  \hline
        \rowcolor{gray!25} & \multicolumn{4}{c}{The 90 $\%$ confidence bound on the absolute error}\\
        \hiderowcolors
        \noalign{\vskip -0.45pt} 
        \hhline{~----}
        \rowcolor{gray!25}\multicolumn{1}{c|}{\multirow{-2}{*}{Material}} &  SLI-EPLD  & Ref. \cite{bizhi}  & Ref. \cite{Chang2023}& Ref. \cite{ITU}\\
        
        \hline
            Glass & 0.0098&0.0419 & 0.0731& 0.0130\\
            \rowcolor{gray!10}Wooden Board & 0.0097 & 0.0514 & 0.0339&0.0137\\
            PVC& 0.0060&0.1138&0.1283&0.0447\\
            \rowcolor{gray!10}Gypsum&0.0114&0.0746 & 0.0782 & 0.0307\\
            Acrylic sheet&0.0129&0.0557&0.0629&0.0219\\
            \rowcolor{gray!10}Tile& 0.0118&0.0692&0.1065&0.0160\\            Concrete&0.0146&0.1233 & 0.1226 & 0.0595\\
            \rowcolor{gray!10}Aluminum& 0.0016&0.0353&0.0328&0.0249\\
            Stainless steel&0.0015&0.0268 & 0.0337 & 0.0212\\
            \hline
	\end{tabular}
        }
\end{table}

The performance differences stem from the models' ability to characterize frequency dispersion. The models in \cite{bizhi} and \cite{Chang2023} neglect parameter variations across wide frequency bands. 
While both our model and the Rec. ITU-R P.2040-4 model \cite{ITU} account for frequency dispersion, our SLI-EPLD model advances further in two key aspects: ($i$) it incorporates a multi-parameter microscopic physical model which outperforms the macroscopic empirical model, and ($ii$) its parameterization strategy maps physical parameters to a stable numerical space, avoiding the extreme dynamic ranges that can hinder robust fitting. This results in a more precise and reliable characterization of the reflection coefficient.

\section{Conclusion}\label{V}
In this letter, we proposed a high-accuracy reflection coefficient model for the THz band, where the WF-TREND algorithm was developed to extract model parameters. The core contribution is the SLI-EPLD model, which has been validated across a diverse set of materials. The experimental measurements and modeling of reflection coefficients in 300$\sim$400 GHz significantly enrich the high-frequency database for RT software. Moreover, the proposed model provides a critical foundation for channel modeling in complex indoor and outdoor environments in the THz band, enabling accurate prediction of signal reflection behaviors. These contributions directly support the development of THz communication systems.

\bibliographystyle{IEEEtran}
\bibliography{reference}

\end{document}